\documentclass[aps,pra,twocolumn,english]{revtex4}

\usepackage{multirow}

\usepackage[T1]{fontenc}
\usepackage[latin9]{inputenc}
\usepackage{babel}
\usepackage{graphicx,amsmath,amssymb,color}
\usepackage[normalem]{ulem}
\usepackage{amsfonts}
\usepackage[toc,page]{appendix}

\usepackage{hyperref}
\usepackage{latexsym}
\usepackage{amsfonts}
\usepackage{algpseudocode}
\usepackage{amsthm}
\usepackage{mathrsfs}
\usepackage{natbib}
\usepackage{color,verbatim}
\usepackage{psfrag}
\usepackage[svgnames]{xcolor}
\usepackage{tikz}

\newcommand{\ket}[1]{|#1 \rangle}

\begin{document}

\title{Fluctuation guided search in quantum annealing}
\author{Nicholas Chancellor}
\email{nicholas.chancellor@gmail.com}
\address{Department of Physics and Durham Newcastle Joint Quantum Centre\\ Durham University, South Road, Durham, UK}

\begin{abstract}
Quantum annealing has great promise in leveraging quantum mechanics to solve combinatorial optimisation problems. However, to realize this promise to it's fullest extent we must appropriately leverage the underlying physics. In this spirit, I examine how the well known tendency of quantum annealers to seek solutions where more quantum fluctuations are allowed can be used to trade off optimality of the solution to a synthetic problem for the ability to have a more flexible solution, where some variables can be changed at little or no cost. I demonstrate this tradeoff experimentally using the reverse annealing feature a D-Wave Systems QPU for both problems composed of all binary variables, and those containing some higher-than-binary discrete variables. I further demonstrate how local controls on the qubits can be used to control the levels of fluctuations and guide the search. I discuss places where leveraging this tradeoff could be practically important, namely in hybrid algorithms where some penalties cannot be directly implemented on the annealer and provide some proof-of-concept evidence of how these algorithms could work.
\end{abstract}

\maketitle


\section*{Introduction and background}

Quantum annealing, in which combinatorial optimisation problems are mapped directly Hamiltonians and solved using sweeps of Hamiltonian parameters, has been a subject of much interest recently. This is in part due to the wide variety of potential applications, in a diverse range of subjects, including for instance air traffic control \cite{Stollenwerk19a}, hydrology \cite{omalley18a}, protein folding \cite{perdomo-ortiz12a}, flight gate assignment \cite{Stollenwerk18a}, finance \cite{marzec16a,Venturelli18a,Orus18a}, and  even quantum field theory \cite{Abel20a,Abel20b}. This subject has further attracted interest because of the experimental maturity of the flux qubit devices produced by D-Wave Systems Inc. which allow for large scale experimentation.

One crucial direction in the growth of flux qubit quantum annealing is an increase in the variety of controls which users can be applied to the experimental quantum annealing process on flux qubit annealers. Traditionally formulated quantum annealing starts from an easy to prepare ground state of a so called driver Hamiltonian and monotonically interpolates the Hamiltonian to a problem Hamiltonian with an unknown ground state. However, major advantages can be gained by using a different control pattern known as reverse annealing, which starts in a state which is a guess for the solution of the optimisation problem, turns on fluctuations, and searches nearby states in Hamming distance by taking advantage of thermal dissipation \cite{reverse_anneling_whitepaper}. Likewise,  controls have been added which allow different qubits to be annealed differently \cite{anneal_offset_report}.

These new features have proven useful in a variety of ways, reverse annealing for instance is motivated by the ability to implement more complex algorithms than traditional forward annealing \cite{chancellor17b}, and these algorithms have shown promising initial experimental results. For example it was shown in \cite{Venturelli18a} that starting from the output of a simple classical algorithm can lead to a large improvement over forward annealing. References \cite{Ottaviani2018a,Golden2020a} showed that iterative methods can help over non-negative matrix factorization. The work in \cite{King2019a} showed experimentally that adding mutation performed using reverse annealing can aid the performance of genetic algorithms. Furthermore the simulation of the celebrated Kosterlitz-Thouless phase transition in \cite{King2018a} would not have been possible without reverse annealing techniques, and reverse annealing has been used to simulate quantum field theories \cite{Abel20b}.

Similarly, anneal offsets have shown promise in synchronizing the freezing of qubits \cite{Lanting17a,anneal_offset_report}. The algorithmic use of these controls was initially motivated by numerical work which demonstrated that locally varying transverse fields can mitigate perturbative anticrossings \cite{Dickson12a}. Fluctuations which can be mitigated using these tools were shown \cite{King16a} to be important in the heavy tails which were observed in early experimental work on D-Wave QPUs \cite{Steiger15a}. The demonstration that anneal offsets can mitigate these effects came later, in \cite{Lanting17a}.

The tendency of quantum fluctuations to lead to uneven sampling of ground state manifolds has traditionally been viewed as a drawback for  quantum annealing \cite{Matsuda09a,Mandra17a,Konz19a}. However, it has been observed that when coupled with classical techniques, this uneven sampling could be a positive feature because the states which quantum annealers find tend to be very different from those found by classical solvers, and therefore could give a more complete picture of the manifold \cite{Zhang17a} if both were used together. In all of this previous work, the uneven sampling was found to be due to quantum, rather than thermal, fluctuations. While thermal fluctuations play an unavoidable role in the experiments reported in this paper, the tendency to favour solutions with more free spins is likely due to the same quantum effects observed in this previous work.

Furthermore, the tendency of the annealer to seek out states with more free spins plays a fundamental role in the effect observed in \cite{King2018a}. The role of quantum fluctuations in causing uneven sampling is also important to graph isomorphism applications which have been proposed and implemented on quantum annealers \cite{Hen12a,Vinci14a,Izquierdo20a} In this paper, I explore a different advantage of this preferential search, the fact that it tends to find states which are flexible in the sense that some variables can be changed at little to no energy cost.

In this paper I experimentally investigate the role which quantum fluctuations can play in the local search which reverse annealing implements. This is done by using specialized Hamiltonians which represent hard problems for the annealer (although not necessarily hard in the computational sense) and have sets of local minima in their energy landscape where fluctuations are enhanced. I also show that the anneal offsets can be used to guide the search by locally enhancing fluctuations on some parts of the system. The technique of locally enhancing fluctuations is reminiscent of the methods proposed in \cite{Chancellor16c}, and provides some experimental validation of these concepts. 

To do this, I construct problems where there is a planted solution which is guaranteed to have the lowest energy, but which does not support strong quantum fluctuations, but fluctuations can be increased by paying an energy cost. I study structures which include both ``gadgets'' which involve binary variables which, in some configurations, can  be flipped with no energy penalty. I also investigate higher than binary variables encoded into ``chains'' which have a ``soft'' configuration where small changes have little or not effect on their energy, but start in a configuration where a small change has a large effect. I also study the effects of anneal offsets which can locally increase or suppress fluctuations.

I use these fluctuations to trade off optimality in solutions for flexibility, in other words find solutions which are a bit less optimal, but for which certain variables can be changed at little or no cost. These variables could either be binary variables directly represented by qubits, or discrete variables which can be encoded using a method I describe later. I argue that this is a property which is likely to be relevant in some real world situations and give a motivational example of how it can be used in a hybrid quantum classical algorithm to find a more optimal solution in the presence of a global penalty function which is not encoded into the annealer.

On the devices studied here (D-Wave 2000Q quantum processing units, QPU), dissipation plays an important (often positive \cite{dickson13a}) role in the annealing process, and the reverse annealing techniques used here fundamentally rely on dissipation. Dissipation can also play a very detrimental role, as pointed out in earlier works such as \cite{Ashhab06a}, however, when techniques like reverse annealing are included, even devices with more limited coherence still present opportunities for quantum advantage \cite{crosson2020prospects}. Recent experiments, have suggested that improvements can be made by reducing the noise on the current D-Wave devices \cite{noise_reduction_spin_glass_whitepaper,Chancellor20a}, however whether these improvements would continue until the noise is reduced to zero is an open question.

The intuition developed here however is likely to carry over into the more coherent protocols proposed in \cite{Perdomo-Ortiz11,Duan2013a,Grass19a}. This is relevant because coherence rates can be improved through a variety of routes, both in supercoducting flux qubit architectures \cite{low_noise_whitepaper,noise_reduction_spin_glass_whitepaper}, and trapped ion quantum annealers \cite{Bernien2017a}. Furthermore, there is significant evidence that in the fully coherent regime fast, but  coherent, quenches known as `diabatic' quantum computing may be a promising path to a quantum advantage \cite{crosson2020prospects}. This is due both to adiabatic mechanisms involving multiple energy levels \cite{Jansen07a,crosson2020prospects}, and mechanisms related to energy transfer \cite{callison19a,Hastings19a,Callison20a}.

Because the experimental details are only likely to be interesting to readers who engage with experiments on the D-Wave devices at a relatively low level, I have reserved a detailed description for the appendix. For the majority of readers who are unlikely to be interested in this level of detail, table \ref{tab:definitions} should suffice to explain the meaning of different parameters and terms.  In section \ref{sec:Results} I give the core experimental results, demonstrating how fluctuations can enable a tradeoff between optimality and flexibility of solutions, as well as how anneal offsets can be used to guide the search by emulating these non-engineered fluctuations. Next in section \ref{sec:flex_examp}, I give a motivational example of how trading off optimality and flexibility can be useful. I then discuss some of the more detailed aspects of the experimental methods and concluded the paper with some discussion.

\begin{table*}[t]
\begin{centering}
\begin{tabular}{|c|c|}
\hline 
term or quantity & definition\tabularnewline
\hline 
\hline 
\multirow{2}{*}{$s^{\star}$} & Parameter controlling range of reverse annealing search, \tabularnewline
 & $s^{\star}=1$ corresponds to no search, $s^{\star}=0$ to forward
annealing\tabularnewline
\hline 
$\delta s$ & Controls offsets on chains or gadgets which enhance ($\delta s<0$)
or supress ($\delta s>0$) fluctuations\tabularnewline
\hline 
chain & Hamiltonian element which encodes a discrete variable\tabularnewline
\hline 
soft chain & Chain in a configuration where more fluctuations are allowed\tabularnewline
\hline 
soft range & Range of values where more fluctuations are allowed\tabularnewline
\hline 
$\mathfrak{s}$(softness parameter) & Parameter controlling level of fluctuations within soft range\tabularnewline
\hline 
domain wall & Feature used to encode the information on the chains (see appendix)\tabularnewline
\hline 
\multirow{2}{*}{gadget (unlocked)} & Hamiltonian element encoding binary variables \tabularnewline
 & with configurations allowing more or fewer fluctuations\tabularnewline
\hline 
free gadget & Gadget in the configuration which allows more fluctuations\tabularnewline
\hline 
locked gadget & Version of gadget element which does not allow for more fluctuations
(used as comparison point)\tabularnewline
\hline 
\end{tabular}
\par\end{centering}
\caption{\label{tab:definitions} Quick reference for terms and quantities used in this paper, full explanation of the experimental setup can be found in the appendix.}

\end{table*}

\section{Results \label{sec:Results}}

In this section I discuss the results of the experiments, which demonstrate how both existing and introduced fluctuations can be used to guide the search which a quantum annealer performs. First I will introduce how the number of free gadgets or soft chains can be controlled by different parameters, such as the value of $s^\star$ and the anneal offsets applied to the chains or gadgets. Measures of the performance of these different control settings will be introduced in Sec.~\ref{sub:cond_prof} and further discussed in Sec.~\ref{sub:off_lock}. A proof-of-principle example for how guided search can be useful will be discussed in Sec.~\ref{sec:flex_examp}.

\begin{figure}
\includegraphics[width=7 cm]{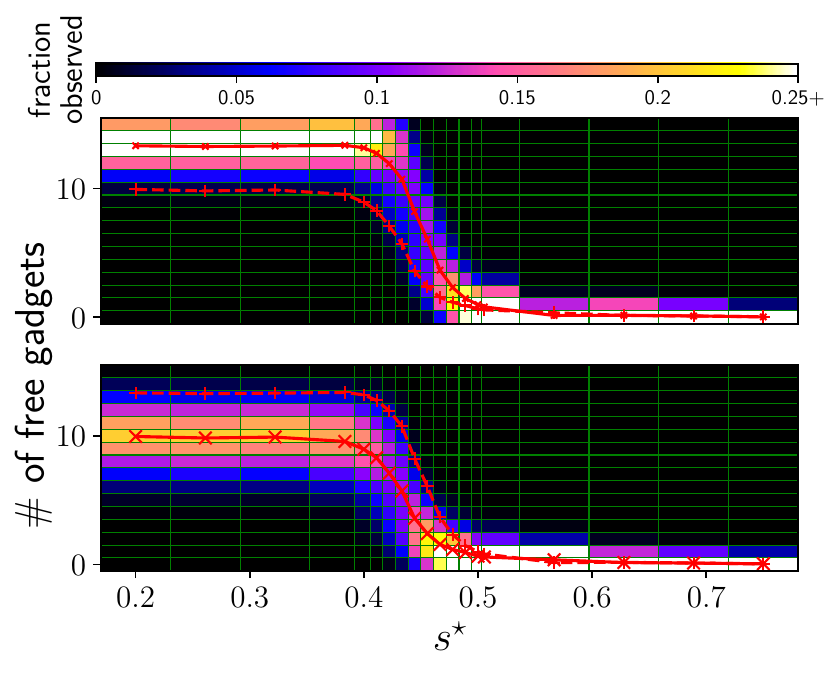}
\caption{\label{fig:free_sprime_gadg_lock_ulock} Fraction of observations for different numbers of free gadgets against different values of $s^\star$ averaged over all $10$ Hamiltonians (recall $s^\star$ is a unitless quantity). Solid red (grey in print) lines with `X' markers represent the mean for the plot, while the dashed line with `+' markers is the mean of the other plot for comparison. Top: Unlocked gadgets with no anneal offsets. Bottom: Locked gadgets with no anneal offsets. The methods for creating the plots given the non-linear mesh are explained in the appendix.}
\end{figure}

The first result which I find is that the number of free gadgets and soft chains both can be increased by decreasing the value of $s^\star$, in other words by increasing the range of the search. Fig.~\ref{fig:free_sprime_gadg_lock_ulock} shows this effect for gadgets, not only are more free gadgets the lower value of $s^\star$, this effect is also much stronger when the gadgets are not locked, indicating that the free variables have a significant effect on the dynamics. For $s^\star\gtrapprox 0.45$ the dynamics are highly localized and very few if any gadgets are free, meanwhile for $s^\star\lessapprox 0.38$, the behaviour is indistinguishable from a search with $s^\star=0.2$, effectively a global search. I have chosen a non-uniform mesh of $s^\star$ values which focuses on the regime where the reverse anneal can lead to long range dynamics, but does not search so far that all information about the initial state is completely forgotten.

The experiments I have conducted can be understood as probing a finite size precursor to a transition from the the phase which is realized at $s^\star=1$ and the paramagnetic phase found at $s^\star=0$. The exact nature of the phase which is realized in the large system limit for $s^\star=1$ is not immediately clear so therefore neither is the nature of the transition and the factors which affect it's location. The fact that annealing from the paramagnetic precursor tends not to find the ground state provides weak evidence against this phase being a simple transition into a ferromagnetic phase, and is more consistent with a spin glass \cite{Vincent18a} or a Griffiths type phase transition \cite{Vojta10a,Nishimura20a}.

\begin{figure}
\includegraphics[width=7 cm]{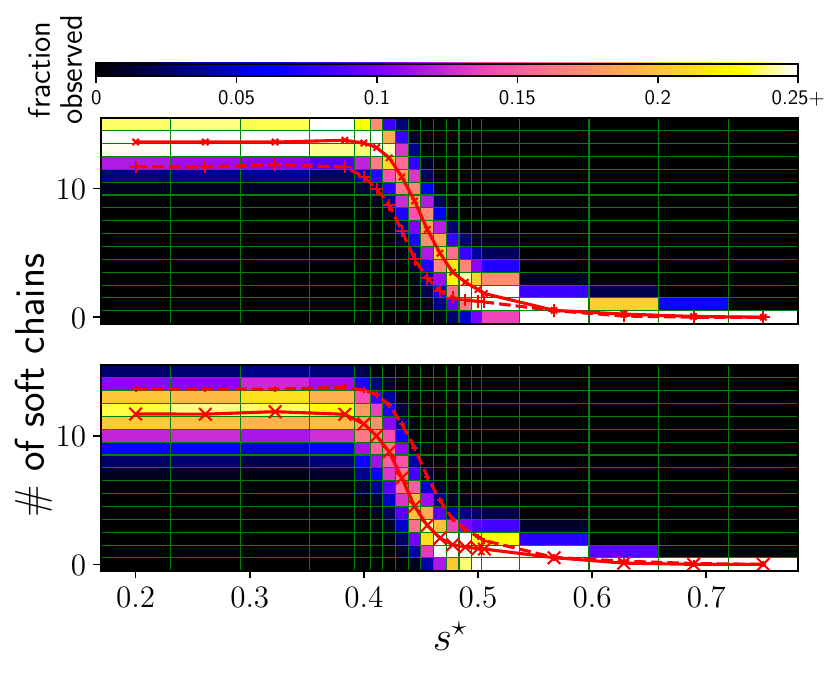}
\caption{\label{fig:soft_sprime_chain_minmax_soft} Fraction of observations for different numbers of soft chains against different values of $s^\star$ averaged over all $10$ Hamiltonians (recall $s^\star$ is a unitless quantity). Solid red (grey in print) lines with `X' markers represent the mean for the plot , while the dashed line with `+' markers is the mean of the other plot for comparison. Top: Minimum softness coefficient chains with no anneal offsets. Bottom: Maximum softness coefficient chains with no anneal offsets. The methods for creating the plots given the non-linear mesh are explained in the appendix.}
\end{figure}

Fig.~\ref{fig:soft_sprime_chain_minmax_soft} shows the same effect for embedded chains in planted solution problems. In this case, reducing the fluctuations by increasing the softness coefficient leads to fewer soft chains. As with the gadget example, non-trivial reverse annealing dynamics are seen for $0.38 \lessapprox s^\star\lessapprox 0.45$. 

\begin{figure}
\includegraphics[width=7 cm]{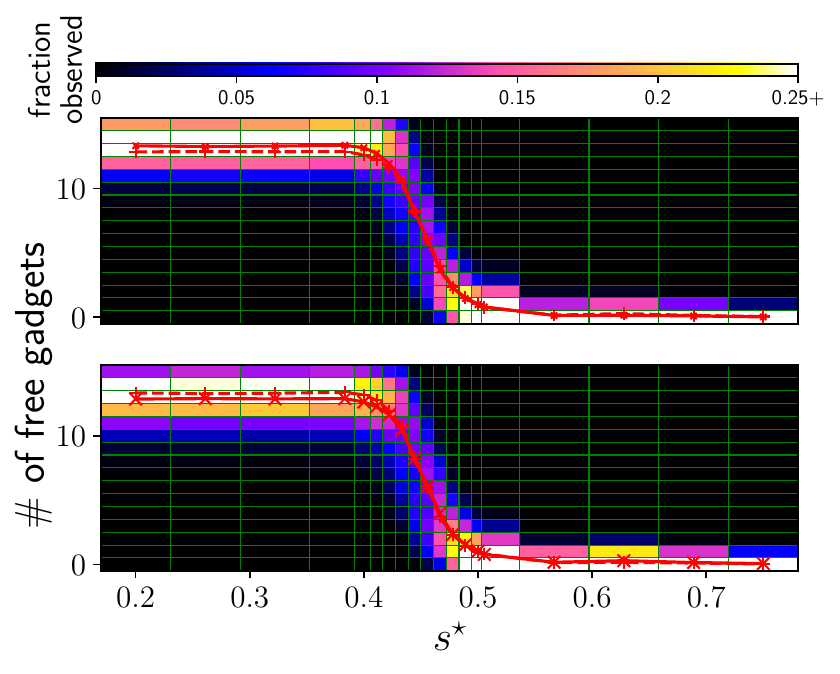}
\caption{\label{fig:free_sprime_gadg_offset_comp} Fraction of observations for different numbers of free gadgets against different values of $s^\star$ averaged over all $10$ Hamiltonians (recall $s^\star$ is a unitless quantity). Solid red (grey in print) lines with `X' markers represent the mean for the plot , while the dashed line with `+' markers is the mean of the other plot for comparison. Top: Unlocked gadgets with no anneal offsets. Bottom: Locked gadgets with anneal offsets ($\delta s$) of up to $-0.04$ applied to the gadgets. The methods for creating the plots given the non-linear mesh are explained in the appendix.}
\end{figure}

We further observe that if we apply anneal offsets to the locked gadgets, we can mimic the effect of the free variables, as Fig.~\ref{fig:free_sprime_gadg_offset_comp} shows the proper choice of anneal offsets renders the distributions indistinguishable for the locked and unlocked gadgets. I show in sec.~\ref{sub:off_lock}, that introduced fluctuations from anneal offsets can be as effective if not more so than fluctuations due to truely free variables.

\begin{figure}
\includegraphics[width=7 cm]{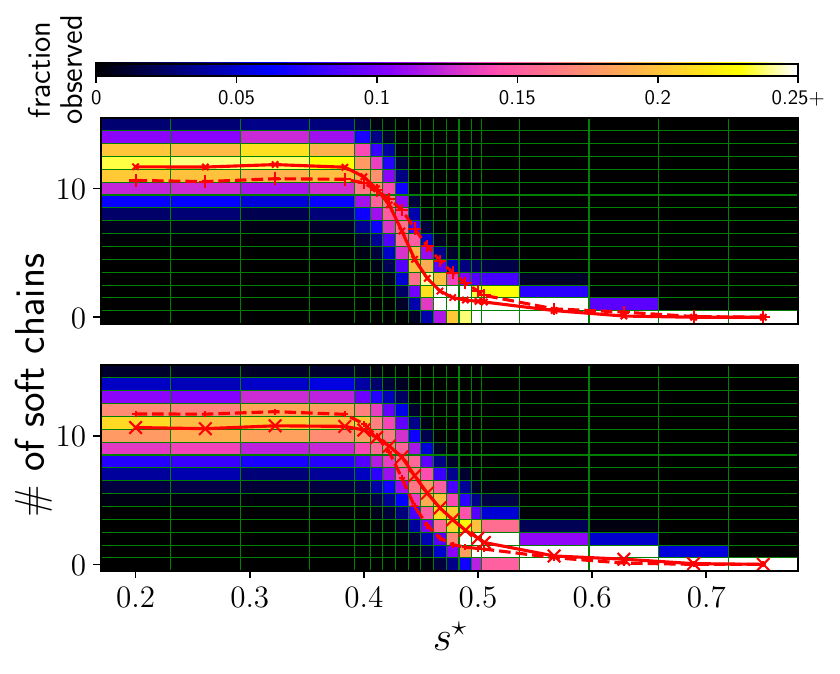}
\caption{\label{fig:soft_sprime_chain_minsoft_offset_effect} Fraction of observations for different numbers of soft chains against different values of $s^\star$ averaged over all $10$ Hamiltonians (recall $s^\star$ is a unitless quantity). Solid red (grey in print) lines with `X' markers represent the mean for the plot , while the dashed line with `+' markers is the mean of the other plot for comparison. Top: Chains with maximum softness coefficient and no anneal offsets. Bottom: Chains with maximum softness coefficient and offsets ($\delta s$) of up to $-0.04$ applied to the gadgets. The methods for creating the plots given the non-linear mesh are explained in the appendix}
\end{figure}

The question now becomes whether anneal offsets can similarly mimic the effect of a lower softness coefficient for chains within the planted solution Hamiltonian. Fig.~\ref{fig:soft_sprime_chain_minsoft_offset_effect} indicates that it cannot, while a negative anneal offset parameter, $\delta s<0$ increases the number of soft chains at intermediate values of $s^\star$, it decreases the number at low $s^\star$. Therefore no value can be used to mimic the behaviour of a lower softness coefficient simultaneously in both regimes. This is likely due to the more complicated structure of the chain encoded discrete variables. I demonstrate in sec.~\ref{sub:off_lock} that in contrast to the locked versus unlocked gadget example, anneal offsets cannot make up the difference between the minimum and maximum values of this parameter.

\subsection{Conditional performance \label{sub:cond_prof}}

Simply analysing solution optimality is a losing proposition, since I have designed the experiments such that, by construction, there is no way of improving beyond the starting condition. However, there is still hope to find high quality solutions which meet conditions which the global solution does not. I define this as \emph{conditional performance}, the best performance attainable which also meets certain conditions. Because of how the gadgets and chains have been constructed, the condition I have chosen to analyse is how many gadgets can be in the free configuration, or chains can be in a soft configuration. This is an interesting criteria since free gadgets and soft chains both make the solution more flexible, allowing for modifications which can be made with little or no energy cost. This flexibility could be important in real world scenarios, for instance if small changes to the solution may need to be made after the time of solving to account for unpredictable events, or if the annealer is being used as part of a hybrid solving technique where difficult to encode global constraints are not included (for an example of the latter see \cite{Venturelli18a}). In Sec.~\ref{sec:flex_examp}, I give an example where flexible solutions can be used to gain an advantage when an additional non-linear constraint is added.

For a fair comparison, we should compare the results from the annealer with a trivial classical strategy of simply frustrating the couplings between the gadgets or chain and the rest of the problem, this `trivial' strategy leads to a cost per gadget or chain of $2$ energy units compared to the most optimal solution. Solutions with a lower cost per gadget/chain, are in principle interesting solutions, whereas those which have a higher energy than the trivial approach are not, since there is a know method which will always attain a better solution using the same starting information. Since the focus of this work is proof-of-concept rather than benchmarking, I will not explore whether or not there are other, less trivial, classical algorithms which can have better conditional performance than the annealer.

\begin{figure}
\includegraphics[width=7 cm]{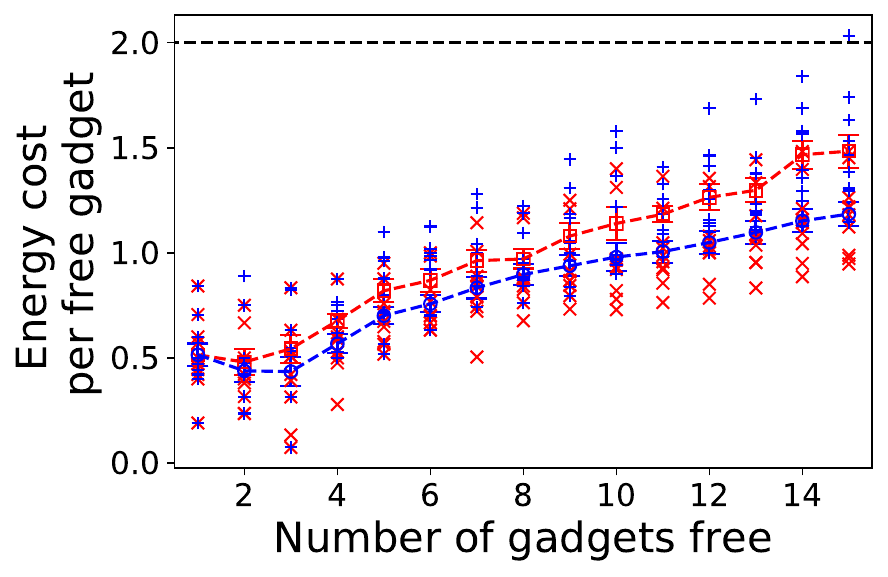}
\caption{\label{fig:cost_number_gadget_off_noOff} Energy cost per free gadget for ten different Hamiltonians using best performing value of $s^\star$ blue plusses  are without anneal offsets, red crosses are best anneal offset (including the possibility of no offset). Red boxes and blue circles represent mean for without and with anneal offsets respectively with error bars representing standard error . Black dashed line is a guide to the eye at a cost of $2$. Energy is in dimensionless coupling units, and $s^\star$ is a unitless quantity.}
\end{figure}

To start off, let us examine the conditional performance for the Hamiltonian with gadgets inserted without using anneal offsets. As Fig.~\ref{fig:cost_number_gadget_off_noOff} shows, even without anneal offsets the annealer is able to outperform a trivial algorithm in all but one case, in which the energy cost is more only if every gadget is made free. When different anneal offsets on the gadgets are allowed, the energy cost per free gadget never exceeds $1.5$.

\begin{figure}
\includegraphics[width=7 cm]{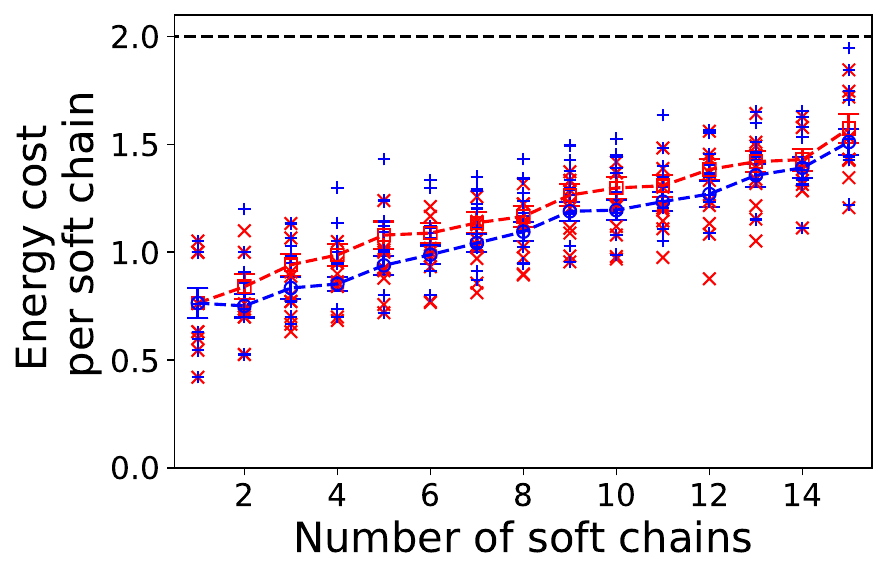}
\caption{\label{fig:cost_number_chainSoft0_off_noOff} Energy cost per soft chain for ten different Hamiltonians using best performing value of $s^\star$ blue plusses  are without anneal offsets, red crosses are best anneal offset (including the possibility of no offset). Red boxes and blue circles represent mean for without and with anneal offsets respectively with error bars representing standard error . Black dashed line is a guide to the eye at a cost of $2$. Softness parameter used was $\mathfrak{s}=0$ in both cases. Energy is in dimensionless coupling units, and $s^\star$ is a unitless quantity.}
\end{figure}

\begin{figure}
\includegraphics[width=7 cm]{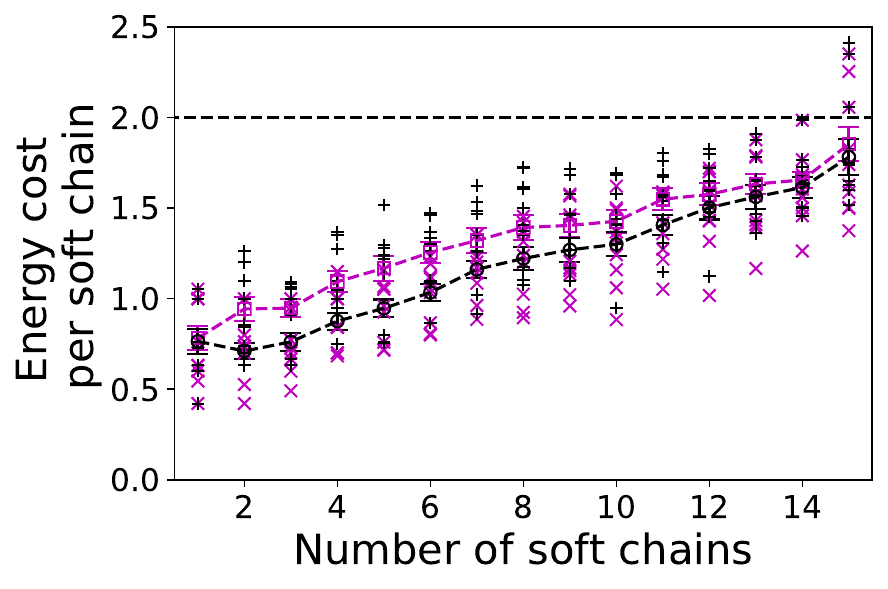}
\caption{\label{fig:cost_number_chainSoft1_off_noOff} Energy cost per soft chain for ten different Hamiltonians using best performing value of $s^\star$ black plusses  are without anneal offsets, magenta crosses are best anneal offset (including the possibility of no offset). Magenta boxes and black circles represent mean for without and with anneal offsets respectively with error bars representing standard error . Black dashed line is a guide to the eye at a cost of $2$. Softness parameter used was $\mathfrak{s}=1$ in both cases. Energy is in dimensionless coupling units.}
\end{figure}

For discrete variables represented as domain walls, reverse annealing is also usually able to find a solution which beats the trivial approach, in fact Fig.~\ref{fig:cost_number_chainSoft0_off_noOff} shows that even without using anneal offsets, the annealer was always able to find a solution which was better than the trivial approach when the soft region of the chain is flat (softness parameter $\mathfrak{s}$ of $0$). Even when the region of the chain which is being searched out is not flat, but a sloping minima (softness parameter $\mathfrak{s}$ of $1$), the annealer is able to beat the trivial approach in most cases, and always does both on average, and for all cases examined with less than $14$ soft chains. The results for the higher softness parameter are depicted in Fig.~\ref{fig:cost_number_chainSoft1_off_noOff}.

I have now shown that reverse annealing in combination with anneal offsets can be effective at modifying solutions to meet certain conditions, but have not elucidated why or how this might happen, in the next subsection I examine potential underlying mechanisms and discuss what the data can teach us about anneal offset strategies.

\subsection{Performance with anneal offsets and locked gadgets\label{sub:off_lock}}

\begin{figure}
\includegraphics[width=7 cm]{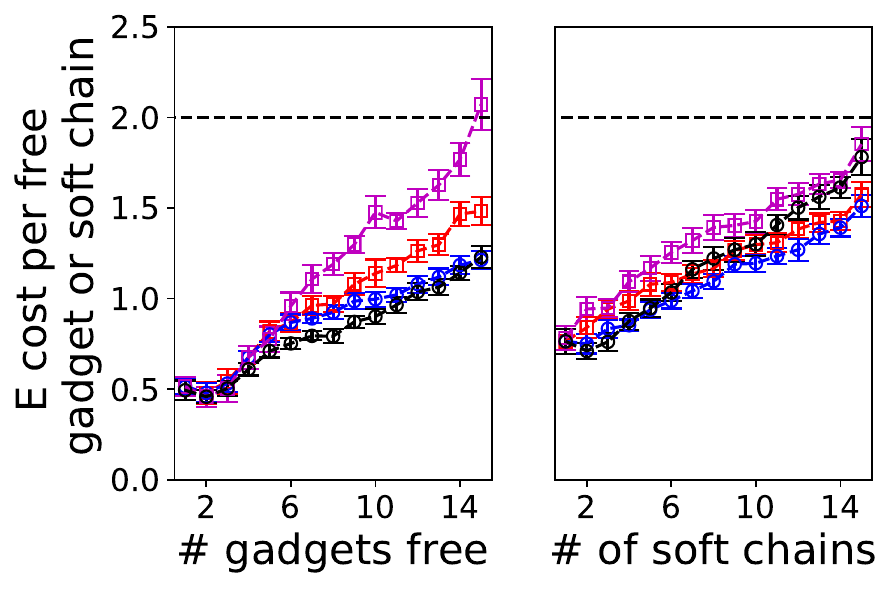}
\caption{\label{fig:cost_compare_chain_gadget} Average energy cost per gadget (left) and chain (right) averaged over ten different Hamiltonians. Magenta boxes and black circles represent mean of best performance without and with anneal offsets respectively and for locked gadgets (left) or softness parameter $1$ (right). Red boxes and blue circles represent the same, but with unlocked gadgets (left) or softness parameter $0$ (right). Error bars represent standard error . Black dashed line is a guide to the eye at a cost of $2$. Energy is in dimensionless coupling units.}
\end{figure}

It is now worth examining more closely the role which quantum fluctuations play in conditional performance, by comparing Fig.~\ref{fig:cost_number_chainSoft0_off_noOff} and \ref{fig:cost_number_chainSoft1_off_noOff} (averages directly compared in Fig.~ \ref{fig:cost_compare_chain_gadget} (left)). We are able to see that better solutions are possible with a lower softness parameter $\mathfrak{s}$, the question we have not explicitly answered yet, is whether the same is true for the fluctuations the free spins cause in the gadgets. To do this we need to compare the `free' and `locked' versions of the gadgets, where free spins are not possible, regardless of the configuration of external spins, which are described in detail in the appendix. As Fig.~\ref{fig:cost_compare_chain_gadget} (right) shows, in the absence of anneal offsets having locked gadgets is very detrimental to performance, at least if more than about $6$ gadgets are desired to be free. On the other hand there is barely any difference once anneal offsets are employed, suggesting that the offsets can enhance the fluctuations and guide the search. Conversely, the effect of anneal offsets seems to be rather minimal for discrete variables encoded in chains.

\begin{figure}
\includegraphics[width=7 cm]{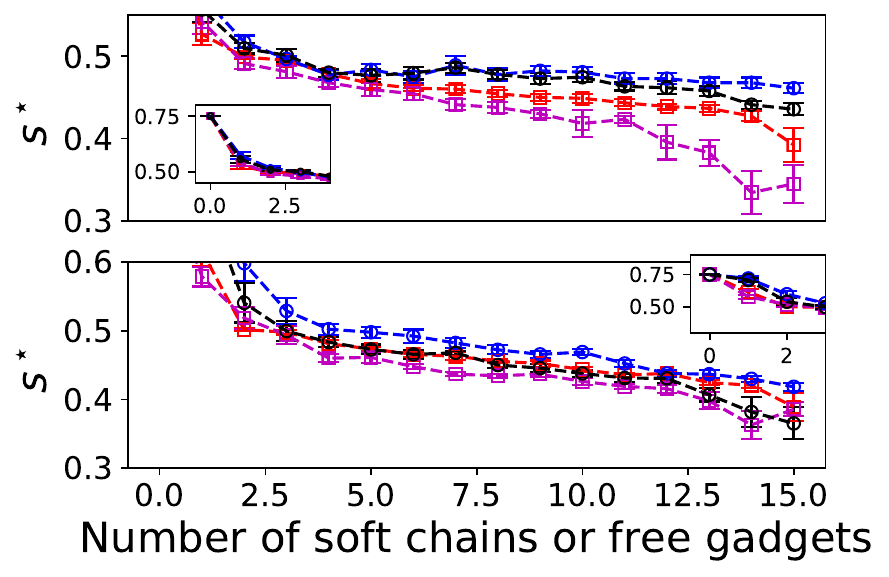}
\caption{\label{fig:s_star_gadg_chain} Average value of $s^\star$ (averaged over ten Hamiltonians) used to obtain optimal conditional performance with a desired number of gagets free, top (soft chains, bottom). Red and magenta squares represent cases where no anneal offsets are used and are unlocked (softness parameter 0) and locked (softness parameter 1) respectively. Blue and black circles represent represent the cases where anneal offsets are used, and and are unlocked (softness parameter 0) and locked (softness parameter 1) respectively. Error bars represent standard error. In all cases the largest value of $s^\star$ was taken in the event of a tie. Insets are the same plots but zoomed out. Energy is in dimensionless coupling units.}
\end{figure}

The first question to ask is what is the optimal value of $s^\star$ for given a desired number of free gadgets and soft chains, and how is this affected by factors like whether or not gadgets are locked and the softness parameter used for chains, as well as whether or not anneal offsets are used. Fig.~\ref{fig:s_star_gadg_chain} shows the optimal value of $s^\star$ for both gadgets and chains under different circumstances. The first thing to notice from this figure is that, perhaps unsurprisingly, $s^\star$ decreases monotonically (within statistical uncertainty) with the desired number of free gadgets or soft chains, this behaviour makes intuitive sense, because changing more variables requires a broader search. Furthermore, constant with Fig.~\ref{fig:cost_compare_chain_gadget}, the values of $s^\star$ based on whether or not anneal offsets are used differ much more for gadgets than for chains, indicating that allowing anneal offsets greatly changes the optimal strategy for gadgets, and does not change it as much for chains. Furthermore,  except for when about $14$ or more free gadgets are desired, the optimal value of $s^\star$ when anneal offsets are used is almost the same for locked and unlocked gadgets, supporting the hypothesis that increased fluctuations from anneal offsets can act as an effective proxy for truly free variables.

\begin{figure}
\includegraphics[width=7 cm]{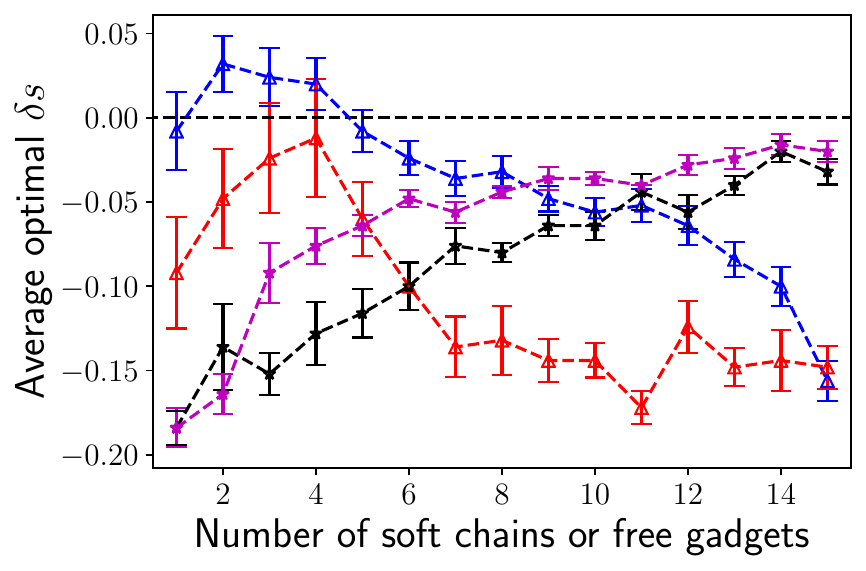}
\caption{\label{fig:optimal_offset_compare} Average anneal offset $\delta s$ taken for $10$ Hamiltonians. Blue (red, lighter grey in print) triangles represent Hamiltonians with unlocked (locked) gadgets. While black (magenta, grey in print) stars represent Hamiltonians with chains with a softness parameter of $0$ ($1$). Error bars represent standard error. Note that positive offsets indicate that fluctuations are suppressed, relative to the rest whereas negative indicates that they are enhanced. In the event of a tie, the lowest numerical value of the offsets which gave the tying energy were taken. Energy is in dimensionless coupling units.}
\end{figure}

To better understand the role anneal offsets are playing, it is worth examining how the best choice of anneal offset depends on the number of free gadgets, or soft chains desired. As Fig.~\ref{fig:optimal_offset_compare} shows, the best strategy is indeed to use stronger offsets in the locked gadget case, and to use them to enhance rather than suppress fluctuations on the gadgets, suggesting that there is indeed a mechanism where offsets artificially guide the search by making the locked gadgets behave as if they have free qubits.  

Fig.~\ref{fig:optimal_offset_compare} further shows that  domain wall encoded discrete variables show very different behaviour to the 
 gadgets, in particular, up to statistical uncertainty, the offsets used in the discrete variable case monotonically approaches zero as more soft chains are desired, while for gadgets with free binary variables, there is non-monotonic behaviour, and a trend toward locally enhancing  fluctuations if more free gadgets are desired. This difference is likely due to the more complex structure of the domain wall encoded variables, leading to less tolerance to fluctuations before they no longer faithfully encode the intended variable.

\section{Motivational example for flexible solutions \label{sec:flex_examp}}

Now that we have shown that the underlying dynamics of quantum annealers can be used to find solutions which are more flexible, it is worth demonstrating an example where such solutions could be useful. To do this, we consider a problem which natively fits onto the chimera graph, but is also subject to  global non-linear penalty. Such global penalties are likely to be encountered in realistic problems, and for example may arise when a shared resource is being used for different purposes and there is a penalty which depends on the \emph{total} amount required. A simple example of how such a constraint could arise in the real world is minimising the total cost of a project if a company owns $X$ number of a piece of equipment, so there is no penalty for a solution which uses any number up to $X$, however there is a cost associated with renting every additional piece of equipment beyond the original $X$. 

While techniques are known to implement global non-linear penalties on quantum annealers, for example those proposed in \cite{chancellor16a,chancellor17a}, these techniques require a fully connected graph and  number of auxilliary qubits equal to the number of original qubits, such an encoding is not practical for large problems on existing quantum annealers. We consider an alternative strategy for solving such problems, we first encode the entire problem except for the global penalty onto the annealer, and use reverse annealing techniques to find solutions with various levels of trade-off between flexibility (for example measured by the number of free gadgets) and optimality. I then perform greedy optimisation as described in the methods section starting from the best solution found at each level of flexibility. This greedy optimisation is performed against the entire problem including the non-linear penalty.

\begin{figure}
\includegraphics[width=7 cm]{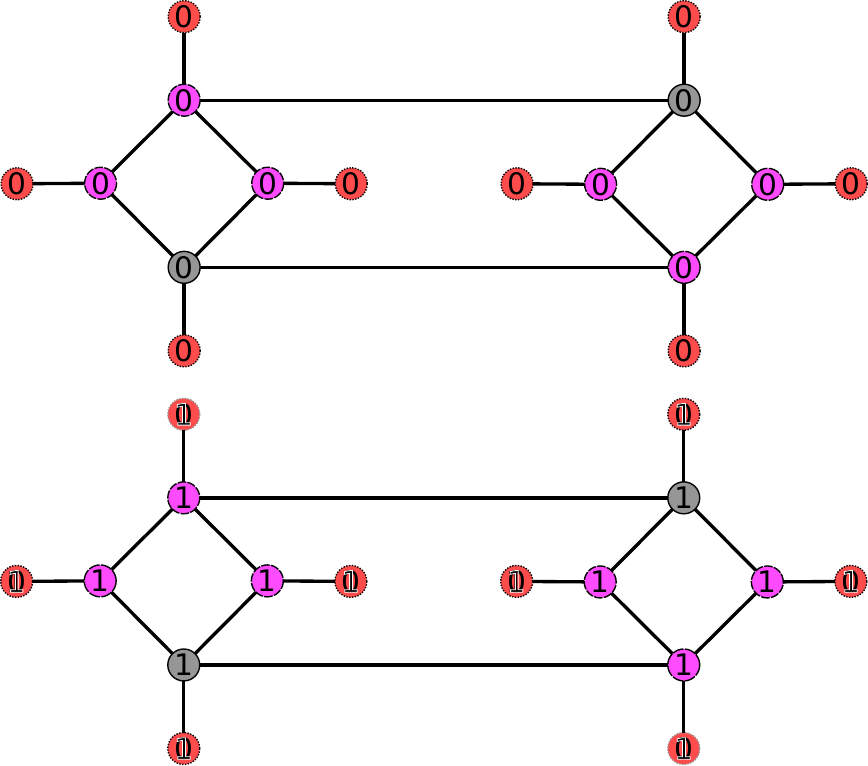}
\caption{\label{fig:therm_asst_ham_gs_fe} The $16$ qubit gadget used in \cite{dickson13a}. Edges represent ferromagnetic coupling of unit strength, and circles represent qubits. Red colouring (and dotted borders) indicates that a qubit is subject to a field of $+1$, while magenta colouring (and dashed borders) indicates $-1$ and grey (solid borders) indicates no field. On the top figure diagram the arrows indicate the unique ground state which satisfies $+1$ fields on the outer qubits, but frustrates the $-1$ field. The bottom diagram is the first excited manifold, where a superimposed $0$ and $1$ indicates that a qubit is ``free'' and can take either value without affecting the energy. }
\end{figure}

Before considering the results for the QPU-sized problems used in earlier demonstrations, it is worth demonstrating this approach with a simpler $16$ qubit example. To do this, we consider the Hamiltonian used in \cite{dickson13a} which is in turn similar to the Hamiltonian considered in \cite{Boixo2013a}. This Hamiltonian has both a local minimum where eight of the $16$ qubits are ``free'', able to exist in either the zero or one state without incurring an energy penalty, and a global minimum where none of the qubits are free, the (unique) ground state and first excited state manifold of this Hamiltonian are depicted in Fig.~\ref{fig:therm_asst_ham_gs_fe}. At least for short runtimes, the close avoided crossings in these devices mean that quantum annealers will typically find the false minimum with more free qubits due to a close avoided crossing relatively late in the annealing schedule \cite{dickson13a}.

We now consider the ability of the solution to adjust to non-linear penalties of different strength. The global non-linear penalty I elect to use is non-linear function of the Hamming distance $\mathfrak{D}$ from a random state
\begin{equation}
E(\mathfrak{D})=1-\exp(\frac{(\mathfrak{D}-(\frac{n}{2}+\sqrt{n+1}))^2}{n+1}),
\end{equation}
where $n$ is the number of qubits involved in the Hamiltonian. The states which the annealer returns will be a Hamming distance $\mathfrak{D}=\frac{n}{2}$ away from most random states, therefore this penalty offsets the Gaussian from the point where a typical solution will sit by its standard deviation, $\sqrt{n+1}$. This will guarantee that the non-linear penalty will have a substantial gradient for typical starting states.

\begin{figure}
\includegraphics[width=7 cm]{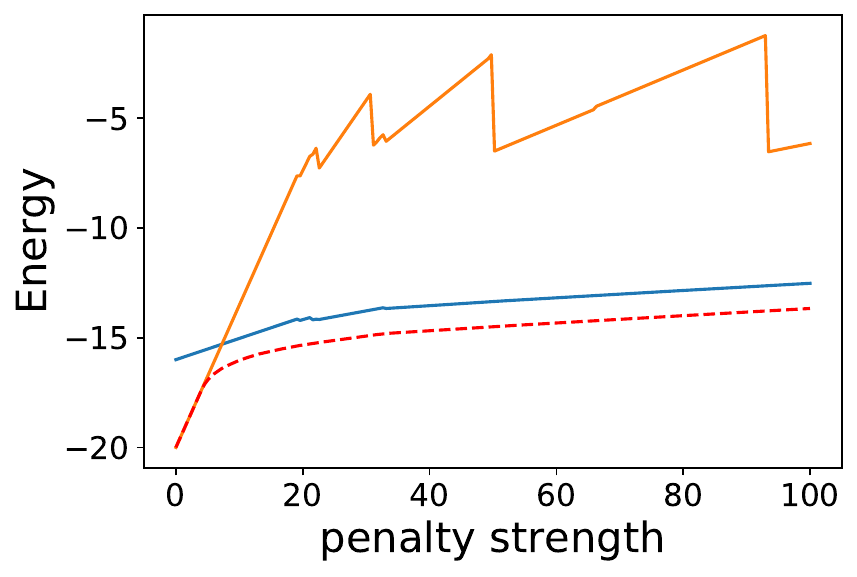}
\caption{\label{fig:16_qb_example} Energy in dimensionless coupling units after applying a non-linear penalty with a strength given by the x-axis and performing greedy search. The gold (light grey in print) line shows results for starting from the true minimum, while the blue (darker grey in print) line shows the result starting from the higher energy, but more flexable false minimum the annealer typically finds. The dashed red line is the energy of the true lowest energy state. $10,000$ samples were taken for each point on this plot, and statistical errorbars are smaller than the depicted lines. }
\end{figure}

Equipped with this definition we consider the results of adding a non-linear penalty followed by a greedy search for the $16$ qubit problem mentioned earlier. As Fig.~\ref{fig:16_qb_example} shows, it is much easier for the greedy search heuristic to compensate for the global non-linear penalty starting from the higher energy but more flexible solution which the annealer finds as compared to the true minimum, the result is that for moderate penalty strength, the more flexible state is a superior choice for a starting configuration.

\subsection{Synthetic use case: optimizing with global non-linear penalties}

\begin{figure}
\includegraphics[width=7 cm]{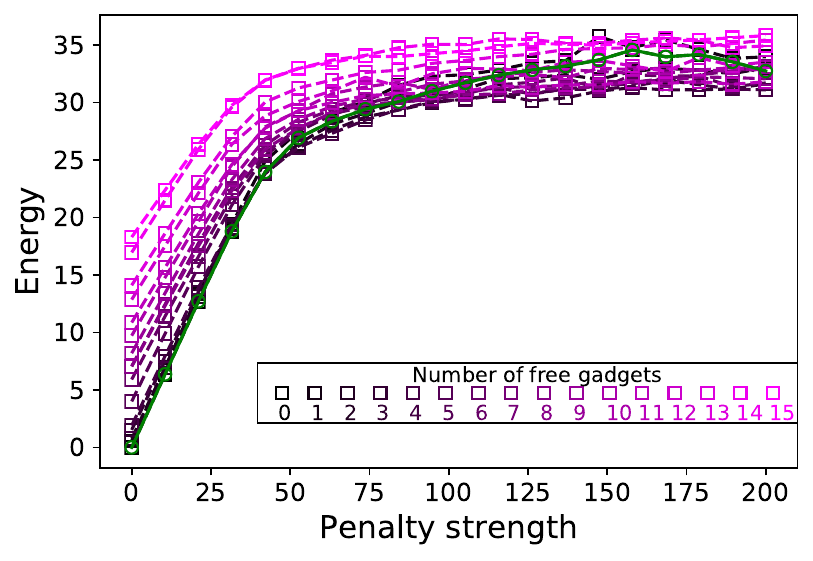}
\caption{\label{fig:large_sys_NLpen_example_Ham7_o} Energy in dimensionless coupling units of the best solution (where planted solution energy is defined to be zero) with a different number of free gadgets versus non-linear penalty strength. The colour encoding of the number of gadgets is depicted in the inset. This plot is for Hamiltonian number $7$ and for the best solution found including the use of anneal offsets, although it is typical of the behaviour seen in both cases. The green (dark grey in print) line is included as a visual aid and follows the state with zero free gadgets. These data were averaged over $300$ choices of random states, and in cases where multiple states were tied for the lowest energy for a given number of free gadgets, a new state was chosen at random for each sample.}
\end{figure}

We now consider what happens when we apply a non-linear penalty followed by greedy search to states with different numbers of free gadgets found for QPU scale problems. While neither the original problem, nor the non-linear penalty are based on anything which one might encounter in the real world, recall that situations where a problem containing a non-linear penalty must be solved are realistic, this can therefore be considered a ``synthetic'' use case for a quantum annealer, not directly based on an application, but with a structure which is likely to be encountered in the real world. We start by considering the best solutions the annealer could find with different free gadget numbers for a single Hamiltonian, in this case Hamiltonian number $7$. As Fig.~\ref{fig:large_sys_NLpen_example_Ham7_o} shows, as the penalty strength is increased to a moderate value, the best solution is no longer obtained from starting a greedy search at the true energy minimum, but from starting with a more flexible state with more free gadgets. For these experiments I only consider the best solution found with each number of free gadgets, choosing at random in the event of a tie. All greedy searches are performed on the unlocked gadgets, where having free variables is likely to improve the solution quality when the non-linear penalty is added.

\begin{figure}
\includegraphics[width=7 cm]{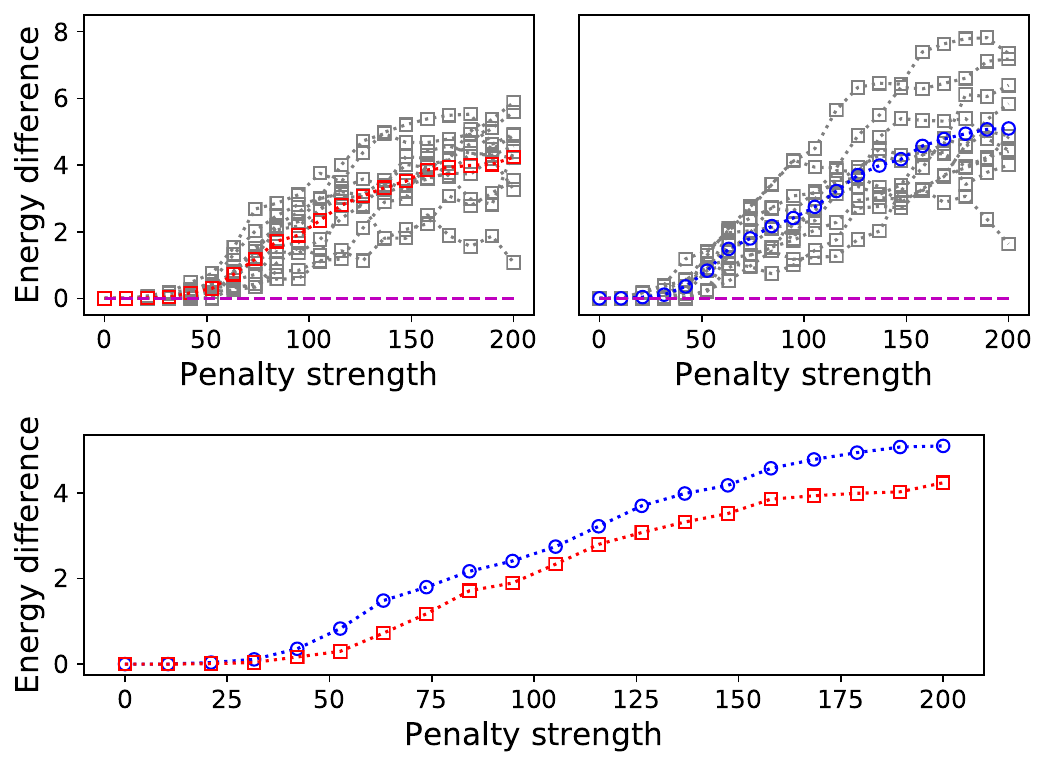}
\caption{\label{fig:E_diff_NLpen} Energy difference in dimensionless coupling units between greedy search performed with a non-linear penalty starting in  planted solution (with no free gadgets) and the best performing state found via reverse annealing. Top right figure only considers solutions found without using anneal offsets, while the top right one is the same but including offsets. The coloured lines represent the average over all ten Hamiltonians while the grey squares represent individual Hamiltonians. The bottom plot shows only the averages, with the blue (dark grey in print) circles representing the method inducing offsets and red (lighter grey than the circles but still darker than background in print) squares without. These data were averaged over $300$ choices of random states, and in cases where multiple states were tied for the lowest energy for a given number of free gadgets, a new state was chosen at random for each sample. Penalty strength is in dimensionless coupling units.}
\end{figure}

From Fig.~\ref{fig:E_diff_NLpen} we can see that the behaviour seen in Fig.~\ref{fig:large_sys_NLpen_example_Ham7_o} is indeed typical of results found both with and without anneal offsets although, unsurprisingly, the cases where anneal offsets are used perform better on average since lower energy solutions can be found by using anneal offsets.

\begin{figure}
\includegraphics[width=7 cm]{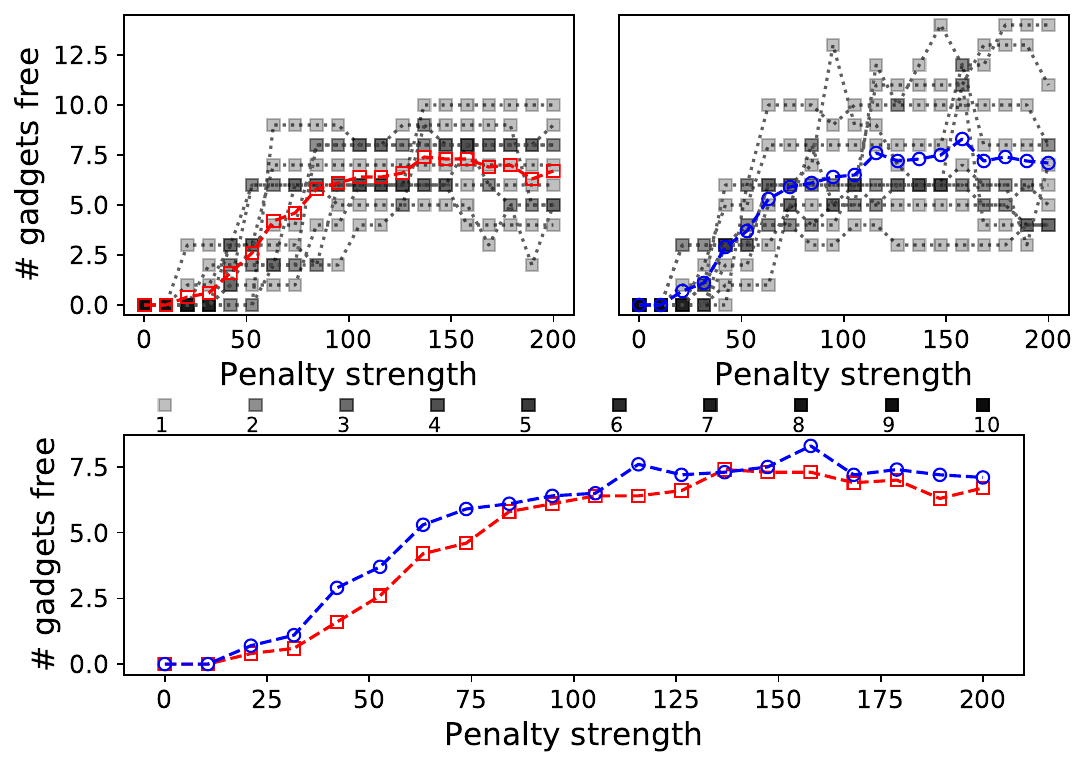}
\caption{\label{fig:opt_num_free_NLpen} Top: number of free gadgets in optimal solution for all ten Hamiltonians versus penalty strength, shading indicates number of Hamiltonians where the same number of free gadgets are optimal for the same penalty strength, a legend for the shading levels appears between the top and bottom figures. The coloured lines are the mean. The left figure is the best solutions not including anneal offsets, while the right is including them. Bottom: The mean from the top two plots shown on the same axis to compare them. Penalty strength is in dimensionless coupling units.}
\end{figure}

Finally, we consider the optimal number of free gadgets in the starting state for different Hamiltonians and penalty strengths. Fig.~\ref{fig:opt_num_free_NLpen} shows that, for both the strategy using anneal offsets, and the one which does not, the typical number of free gadgets in the best performing state increases for a while with penalty strength and then settles to an average across all Hamiltonians of around seven free gadgets. While it is possible that the average number of free gadgets is slightly higher for the strategy using offsets, the difference is relatively small. It is however clear that for the solutions which used anneal offsets, there is a much wider variety of solutions, and in particular, a tendency to use some solutions with many more free gadgets. 

Recall that the gadgets and chains were observed to behave qualitatively differently in the previous section, so the results from gadgets may or may not carry over to chains. The purpose of this section is to provide proof-of-concept for the usefulness of fluctuation guided behaviours, not to provide exhaustive evidence of how they can be used, so I will not analyse these in detail. The data associated with this paper however are publicly available \cite{guided_search_data}, and analysis of the chain data from the perspective of hybrid algorithms, similar to what I have done here for gadgets  is likely to produce interesting results.

\section{Methods \label{sec:methods}}

All reverse annealing experiments were performed using the maximum allowed annealing rate on both the forward and reverse anneal, at this rate the entire (forward) anneal would be completed in $5\,\mu$s. All experiments used a hold time $\tau$ of $20\,\mu$s. All annealer calls were set to perform $1,000$ individual runs. The reverse annealing experiments presented here were performed using the D-Wave Matlab API between 27 October 2018 and 30 October 2018 on a commercially available D-Wave 2000Q QPU with QPU time purchased by BP plc. Data are publicly available at \cite{guided_search_data}.

Greedy optimisation was performed by checking all single bit flips and performing the one which reduces the energy the most, choosing at random in the event of a tie. The greedy procedure is repeated until no single bit flip will reduce the energy.

All plots were produced in the Python language \cite{van2003python} and the matplotlib plotting package \cite{hunter2007matplotlib}, code used to produce the plots and perform the experiments is available from the same public repository as the experimental data. Heat-map plots with non-linear grids were plotted such that the centre of each cell aligns with the value of each axis. The NumPy \cite{oliphant2006guide} and SciPy \cite{scipy} packages were also used as well as jupyter notebooks \cite{jupyter} and the IPython interpreter \cite{perez2007ipython}.

\section{Discussion and conclusions}

In this paper, I have demonstrated how fluctuations can guide quantum annealers to trade off optimality for more flexible solutions, as well as motivated cases where such a tradeoff could be useful. The particular useful case I focus on is when a problem involved global penalties which cannot practically be implemented on the annealer. While in the past the tendency of quantum annealers to find solutions where fluctuations are stronger has been seen as a weakness, for instance in inhibiting the ability to uniformly sample ground states, I demonstrate ways in which it could be useful. 

In addition to demonstrating that the existing fluctuations on the annealer can help guide searches toward more flexible states, I show that locally offsetting the annealing schedule of the qubits can be used to guide the search. This provides experimental motivation for methods like those proposed in \cite{Chancellor16c}, which incorporate bitwise uncertainty into algorithms. 

Realistic problems may have chain like structures if the encode discrete variables, but will not contain gadgets like the ones discussed here. Even for problems which encode discrete variables, it will not necessarily be obvious which chains should have the offsets applied. Identifying areas where there is more potential for flexibility to be increased using anneal offsets in the setting of either binary or higher-than-binary discrete variables is an interesting problem, but one which is beyond the scope of the current work. Techniques could include experimental analysis to see where soft chains or free qubit variables appear and offset those parts of the problem Hamiltonian, but testing would need to be done to determine whether this approach is useful or practical.

While not explored here, it is likely that analogous effects could be seen in quantum inspired algorithms based on spin like systems, for example quantum Monte Carlo techniques \cite{Martonak02a} which should show analogous effects to the fluctuations observed here. In fact the proof-of-concept numerics in \cite{Chancellor16c}, exhibited that fluctuations can attract quantum Monte Carlo dynamics preferentially to some minima over others. This work has introduced new ways in which quantum annealers and related algorithms can be used, beyond directly finding the most optimal solution, an important direction in hybrid quantum-classical computing. By laying the groundwork for how modifying fluctuations locally can be used algorithmically to guide a search, the work here opens a new path to using these modified fluctuation strengths algorithmically, in a similar vein to currently used reverse annealing techniques, but guiding the direction of the search, rather than the starting point.
 
\section*{Acknowledgments}

This work was supported by BP plc, by EPSRC fellowship EP/S00114X/1, EPSRC grant EP/L022303/1 and ESPRC Hub EP/M013243/1. The author thanks Adam Callison for pointing out several typos in the original ar$\chi$iv version, as well as Andrew King for providing key references and perspective for the discussion of previous work.

\section*{Appendix: Experimental setup}

These experiments involve both specially engineered Hamiltonians to construct a search space with the necessary properties, and the use of advanced control features of the QPU, both anneal offsets and reverse annealing, which are used in combination. I first describe how the Hamiltonians are constructed, and than how they are used in the actual experimental protocols. Before, I do this, it is useful to provide some background on the operation of the quantum annealer. In the simplest implementation of it's protocol, QPU realizes a transverse field Ising Hamiltonian
\begin{equation}
H=-A(s)\sum_iX_i+B(s) H_{\mathrm{prob}}, \label{eq:H_tf_no_offset}
\end{equation}
where $A(s)$ and $B(s)$ are functions of a control parameter $0\le s \le 1$, which are non-linear in general, but could also be linear in principle, $X_i$ is a Pauli $X$ acting on qubit $i$, and $H_{\mathrm{prob}}$ is a programmable Ising problem Hamiltonian,
 \begin{equation}
H_{\mathrm{prob}}=\sum_{ij}J_{ij}Z_iZ_j+\sum_ih_iZ_i, \label{eq:H_prob}
\end{equation}
where $Z_i$ is a Pauli $Z$ acting on qubit $i$, the details of how $J_{ij}$ and $h_i$ are chosen is discussed later. 

The QPU is designed so that $|\frac{A(0)}{B(0)}|\gg 1$ and $|\frac{A(1)}{B(1)}|\ll 1$, the ratio $|\frac{A(s)}{B(s)}|$ decrease monotonically with $s$ and neither $A$ nor $B$ change sign during the protocol. I do employ a more advanced feature known as anneal offsets, which slightly changes the form of Eq.~\ref{eq:H_tf_no_offset}, and will be discussed in due course. 

\subsection{Hamiltonian construction}
The goal of the experiments in this paper are to study the ability of a quantum annealer to use fluctuations to find high quality solutions which are flexible in the sense that changing some elements of the solution will not affect the energy of the solution, or will only affect it very little. Since I am not developing this study as a benchmark against classical methods, I have focused on designing Hamiltonians which are difficult to solve for the annealer, and have a known solution, but which are not necessarily computationally hard problems. To this end, the problems used here build on the planted solution construction from \cite{Hen15a}, which yields limited computational hardness \cite{king15a,Hamze18a} (for state-of-the-art solution planting techniques, see \cite{Perera20a}). Furthermore, I use many more clauses than would be desirable to construct the hardest problems in the interest of ensuring that the problem graph is connected and to reduce the degeneracy of the ground state manifold. 

\begin{figure}
\includegraphics[width=7 cm]{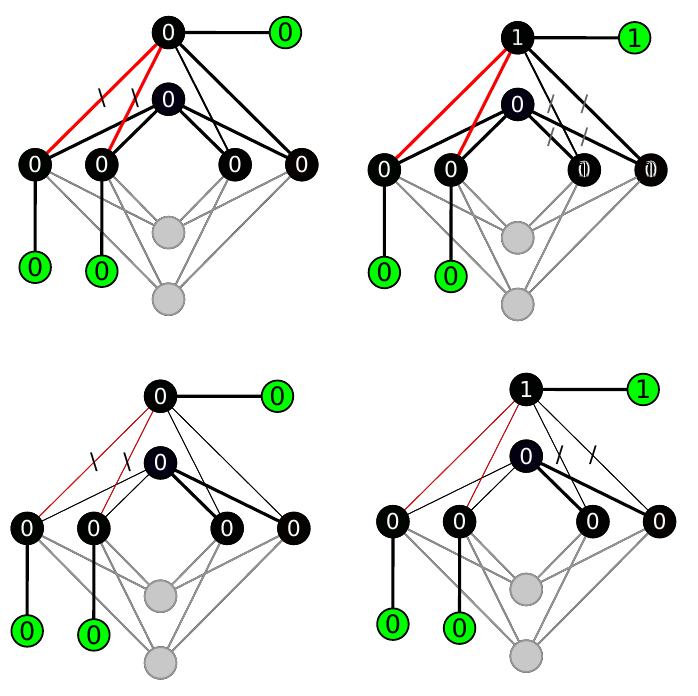}
\caption{\label{fig:free_and_locked_gadget_GS}Top: Minimum energy state of free variable gadget with different configurations of external variables, left is where all external variables agree, right is where one disagrees. Bottom: Same but for locked gadget. Grey edges and circles indicate unused couplers and qubits, green indicates (dark grey in print) external qubits, red (dark grey, between pairs of dark nodes in print) and black edges indicate anti-ferromagnetic and ferromagnetic coupling respectively, while thick edges indicates coupling of unit strength and thin indicate coupling with a strength of $0.5$. Superimposed 1 and 0 characters indicate free variables. Slashes indicate frustrated couplings, with grey slashes indicating multiple possibilities depending on the values of the free variables.}
\end{figure}

\begin{figure}
\includegraphics[width=7 cm]{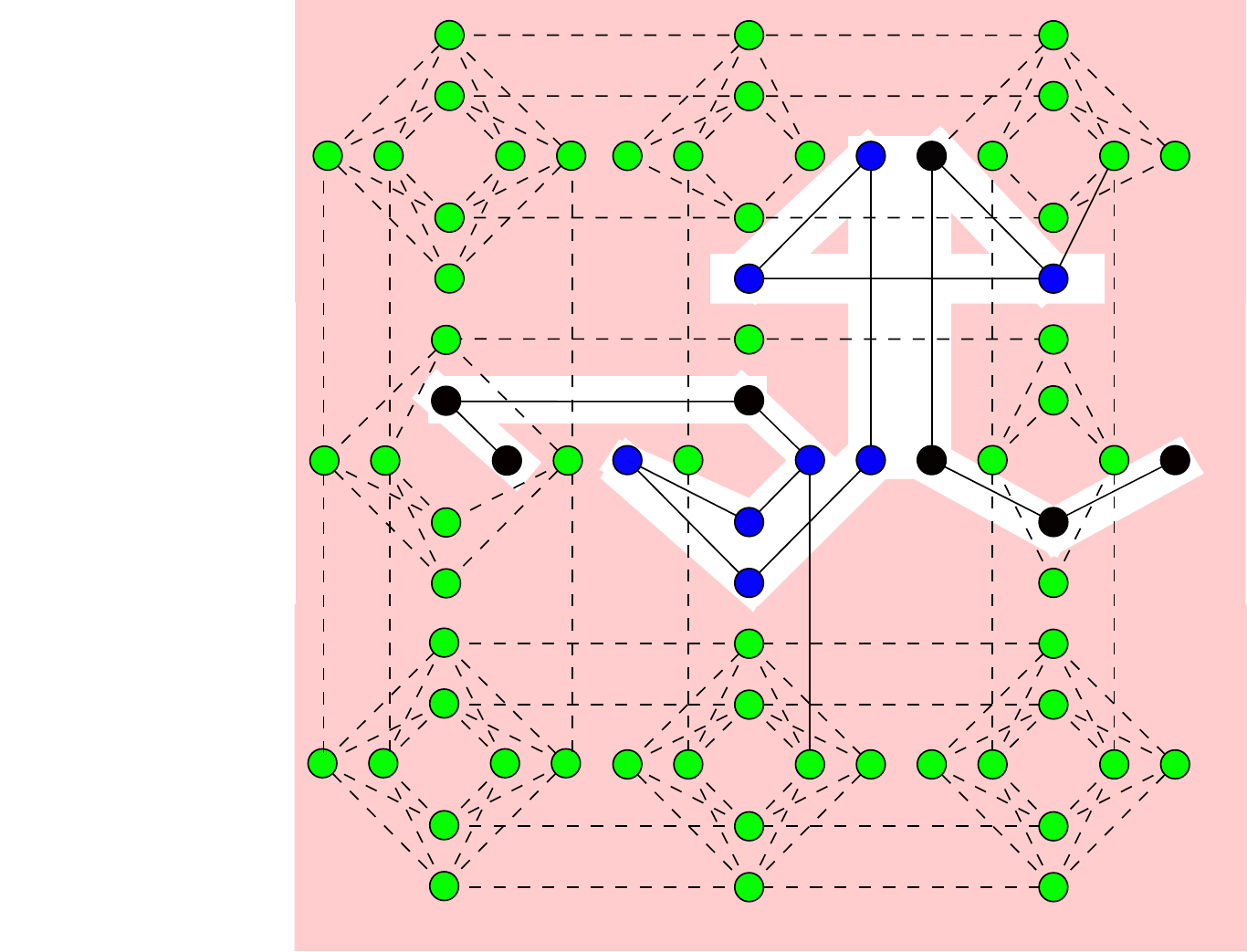}
\caption{\label{fig:chimera_chain_insert} Chain encoding domain wall variable placed within a planted solution problem. Black circles indicate qubits which are part of the domain wall encoding but not within the soft range, blue (darker grey in print) colouring is a guide to the eye to indicate the soft range and the externally coupled qubits at each end of the range. Green (lighter grey in print) indicate those outside of the domain wall variable. Solid edges indicate ferromagnetic coupling of unit strength, while dashed edges indicate couplings which are part of the planted solution encoding. Pink (grey in print) colouring is used as a guide to the eye.}
\end{figure}

The methods which I use, proposed in \cite{Hen15a}, constructs problems with planted solutions by generating overlapped frustrated loops on the edges of the underlying graph via random walks which terminate when they intersect their own path. I use planted solution problems with loop size less than six and $8,000$ loops on a QPU with approximately $2,000$ qubits (some of which are reserved for specialized features as discussed later in this section), with a coupling arrangement knows as a chimera graph. The details of this coupling graph are not important to understand this study, but a full description of the graph can be found here \cite{chimera_describe} Figs.~\ref{fig:free_and_locked_gadget_GS} and \ref{fig:chimera_chain_insert} depict chimera graphs with a $3$x$3$ grid of eight qubit unit cells, the 2000Q has the same eight qubit unit cells arranged in a $16$x$16$ grid.

In addition to having a known planted solution, the experimental Hamiltonians also need features which can explore the ability of the annealer to use fluctuations to find more flexible solutions. Since I intend to study the ability to find more flexible solutions in both a binary and discrete setting in other words both in the setting where a variable can take two values, and the setting where it still takes a finite number of discrete values, but can take more than two, two different strategies need to be employed, gadgets where variables are allowed to become ``free'' should be embedded, henceforth referred to as ``gadgets'' as well as chains of qubits which encode discrete variables using the domain wall encoding described in \cite{Chancellor19a}, henceforth referred to as ``chains''.

\begin{figure}
\includegraphics[width=7 cm]{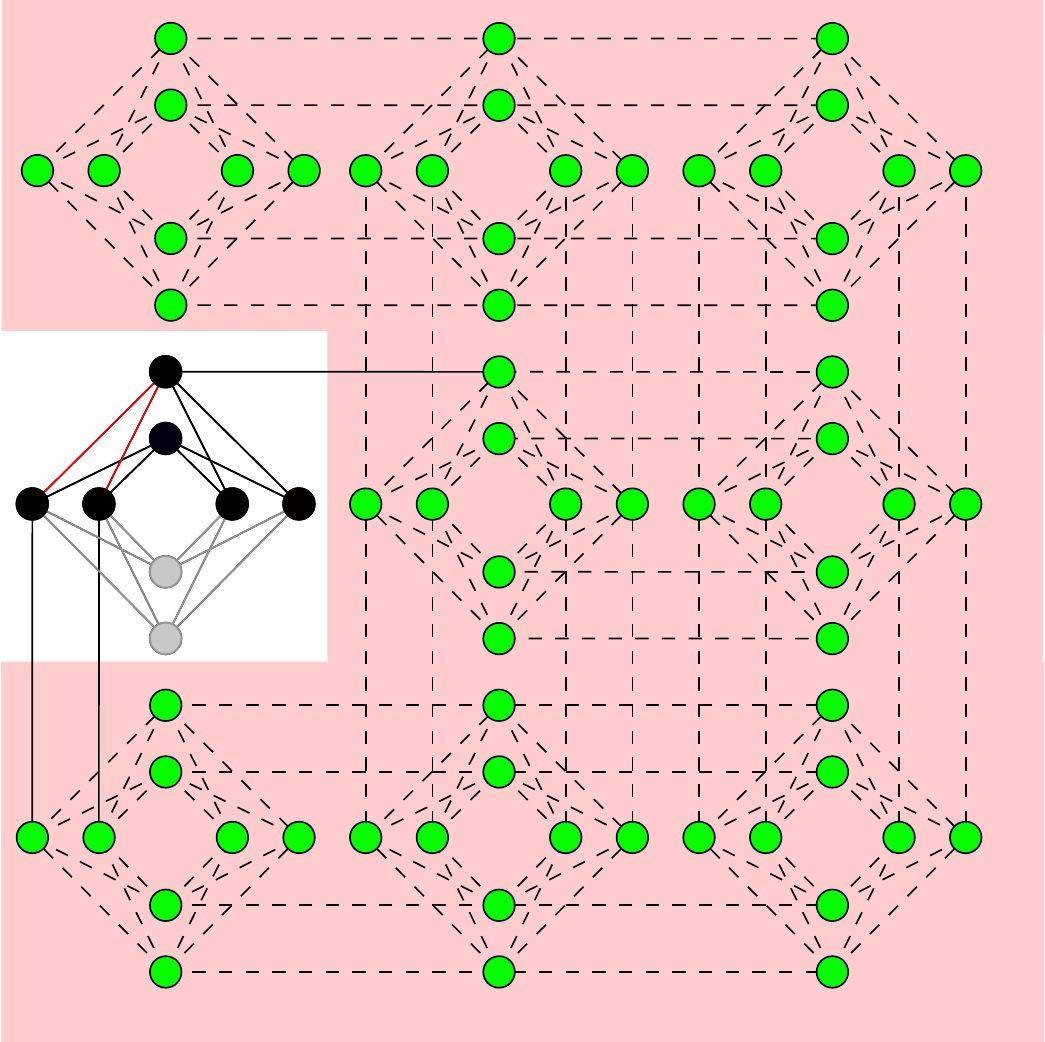}
\caption{\label{fig:chimera_gadget_insert} A free spin gadget inserted into a planted solution Hamiltonian. Green circles (dark grey in print) and dashed edges indicate qubits which are not part of the gadget, while black qubits indicate the active qubits within the gadget, with black edges indicating ferromagnetic couplers of unit strength and red edges indicating anti-ferromagnetic couplers of unit strength. Grey circles and edges indicate the unused couplers and qubits within the gadget. Pink (grey in print) colouring is used as a guide to the eye.}
\end{figure}

Fortunately, the planted solution construction does not require a  full chimera graph to be effective. This means that the construction can be performed with some qubits reserved for either gadgets or chains, and these features can be added in later. The gadgets are constructed with the following properties:
\begin{enumerate}
\item Couplings to neighbouring qubits within the planted solution construction
\item A unique ground state when all external qubits which the gadget couples to are in the $\ket{0}$ (or $\ket{1}$) configuration 
\item Degenerate ground state with free variables in cases where the external couplings do not agree, and therefore the planted solution components may be frustrated, these have equal energy to the unique state
\item Occupy a single chimera unit cell
\end{enumerate}
Fig.~\ref{fig:chimera_gadget_insert} depicts a gadget which obeys these properties embedded into a larger problem Hamiltonian. The top row of Fig.~\ref{fig:free_and_locked_gadget_GS} depicts the lowest energy states of this gadget when external qubits either agree of disagree. To be able to separate the effects of fluctuations due to free spins from other effects, I have also developed a `locked' version of the free variable gadget, in this version some of the couplings are reduced to half the strength of the others so that the lowest energy state when the external variables do not agree no longer contains free variables, these are depicted on the bottom row of Fig.~\ref{fig:free_and_locked_gadget_GS}.

To embed discrete variables, I use the domain wall encoding from \cite{Chancellor19a} to encode a variable with $16$ possible values within a $15$ qubit chain with unit ferromagnetic coupling strength. For completeness, I review the domain wall encoding at the end of the appendix. I use the field controls of the annealer to control the potential on this chain such that the $0$ value of the variable (all qubits are in the $\ket{0}$ configuration), has the same energy as the minimum energy in a `soft' region which corresponds to seven consecutive values of the discrete variable which are randomly chosen to start anywhere from two to six (recall that I use a convention where the allowed values run from $0$ to $15$). The chain is coupled to to the rest of the problem Hamiltonian on the first and last qubit of the soft region, such that the planted solution must be frustrated if the domain wall is in the soft region. All other values of this variable have an energy which is two energy units higher than either the minimum of the soft range or the $0$ state of the variable. Henceforth I refer to a chain where the domain wall is in the soft region as a `soft chain'. The  qubit chain used to encode the domain wall variable is randomly placed within the planted solution problem by performing $15$ steps of a non-self-intersecting random walk on the hardware graph, an example of a chain within the larger Hamiltonian is depicted in Fig.~\ref{fig:chimera_chain_insert}.

The potential within the soft range is always equal to the $0$ value of the variable at the midpoint $m$ of the range. Away from the midpoint the potential increases such that $E(m+j)=E(m)+\mathfrak{s}|j/2|$ where the parameter $0\le \mathfrak{s} \le 1$ is the ``softness parameter'' of the chain. Lower values of $\mathfrak{s}$ allow for more fluctuations since it costs less energy for the domain wall to move away from the centre of the chain.

For both the gadget and chain versions of the problem, ten Hamiltonians were created at random. Other than the reduced strength of the gadget couplings, the free and locked gadget runs use the same ten Hamiltonians. Each Hamiltonian incudes either 15 gadgets or 15 chains.

\subsection{Annealing protocol}

\begin{figure}
\includegraphics[width=7 cm]{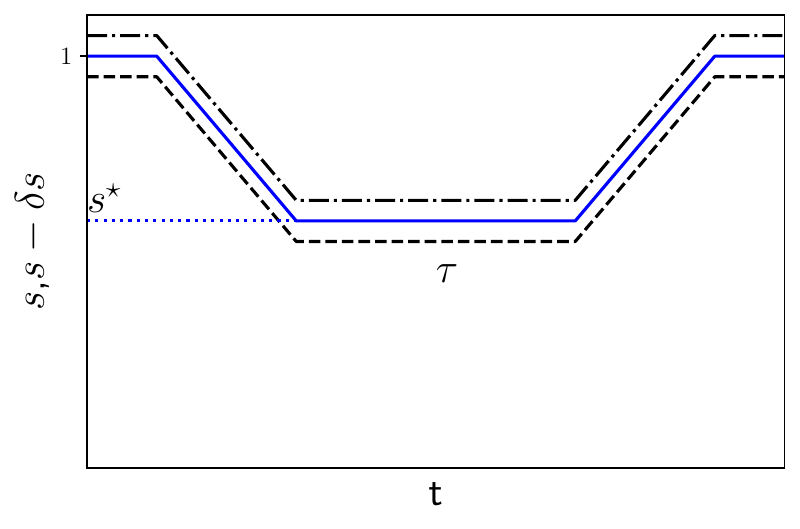} \caption{\label{fig:rev_anneal_offsets} Schematic of reverse annealing protocol, solid line is the value of $s$ for qubits not in gadgets or chains (or all qubits in no offset case). Dot-dashed line is the schedule for chains and gadgets with positive $\delta s$, the dashed line is the same for negative $\delta s$. The quatity $s$ is unit less and time is typically in units of $\mu$s}
\end{figure}

The key feature of the Hamiltonians constructed for these experiments is that they have known planted solutions. This is crucial for the purpose of this study, to explore the ability of the device to trade off between optimality and flexibility, we need to start off in a state which is known to be optimal. Fortunately the reverse annealing feature \cite{reverse_anneling_whitepaper} allows for a search around the planted (or any other classical) state. The reverse annealing feature uses a protocol which starts the QPU in a state determined by the user at $s=1$, anneals to a value $s^\star$ held for a time $\tau$ and then anneals back to $s=1$ as depicted in Fig.~\ref{fig:rev_anneal_offsets} . Thermal dissipation allows the device to seek out lower energy states during the reverse annealing protocol.

In addition to the reverse annealing feature, I also make use of another feature called anneal offsets \cite{anneal_offset_report}. The function of this feature is to offset the annealing parameter on different qubits. In particular, I offset the parameter values of either the chains or the gadgets (a subset of qubits I call $g$), which makes the Hamiltonian
\begin{align}
H=-A(s)\sum_{i\notin g}X_i+ \nonumber\\
B(s)(\sum_{i\notin g,j\notin g}J_{ij}Z_iZ_j+\sum_{i\notin g}h_i)-A(s+\delta s)\sum_{i\in g}X_i+\nonumber\\
B(s+\delta s)(\sum_{i\in g,j \in g,j \notin g}J_{ij}Z_iZ_j
+\sum_{i\in g}h_i), \label{eq:H_tf_offset}
\end{align}
where $i \in g$ means that qubit $i$ belongs to a gadget or chain and $i \notin g$ means that it does not. Effectively, for a positive value of $\delta s$ the strength of the couplings and longitudinal field terms within $g$ as well as coupling between $g$ and the rest of the qubits is increased, while the transverse field within $g$ is decreased. For negative $g$ the opposite happens. In simplified terms, positive $\delta s$ leads to weaker fluctuations with $g$, while negative $\delta s$ leads to stronger. Since the original purpose of the anneal offset purpose was to suppress fluctuations, previous work has focused most strongly on positive $\delta s$.

\begin{figure}
\includegraphics[width=7 cm]{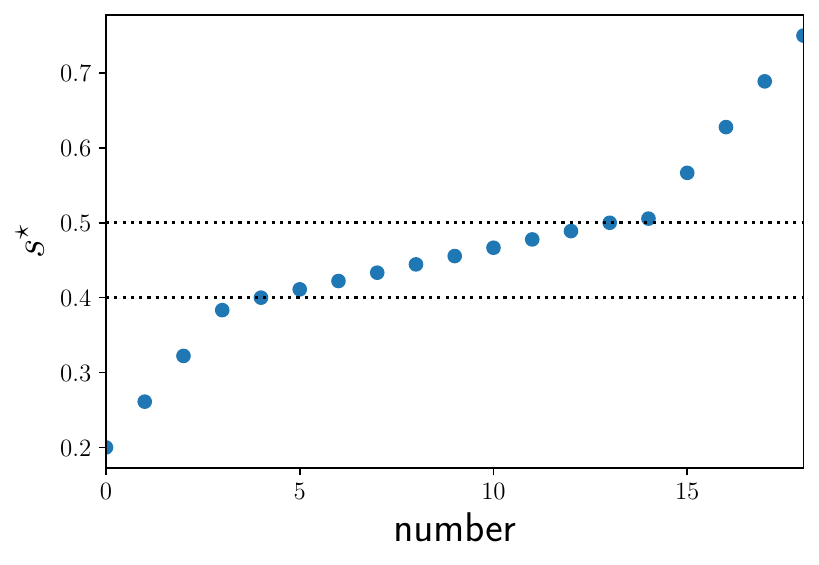} \caption{\label{fig:s_star_grid} Grid of values of $s^\star$ used in this study, dotted lines are guides to the eyes to show the region of higher interest, between about $s^\star=0.4$ and $s^\star=0.5$. Both quantities in this plot are unitless.}
\end{figure}

For all experiments reported here, the anneals in the reverse annealing protocol were performed at the maximum allowed rate, which traverses from $s=0$ to $s=1$ in $5\,\mu$s and a hold time $\tau$ of $20\,\mu$s was used. As a comparison point, the energy scale of these devices lead to natural frequencies in the GHz range. The same parameters were used for chains and gadgets. For all values of $s^\star$, I used a linearly spaced grid of $11$ values of $\delta s$ evenly spaced between $-0.2$ to $0.2$, inclusive of the end points. Since not all qubits are capable of the full range of offset values, the maximum magnitude allowed (positive or negative) value was used when the desired value fell outside of the range. Because I wanted to study extreme values of $s^\star$ as well as studying more values within a region of interest where the data were observed to change rapidly with $s^\star$, I chose a the non-uniform grid of $19$ values of $s^\star$ depicted in Fig.~\ref{fig:s_star_grid}.

\subsection{Review of the domain wall variable encoding strategy}

The domain wall encoding strategy used here was originally developed to undertake the research described in this paper, since it can encode discrete variables on a chimera graph without requiring minor embedding unlike the more traditional one-hot strategy. Because the domain wall encoding strategy has been observed to significantly out perform the one-hot strategy on several key metrics related to embedding on realistic hardware graphs, a full description of this technique as well as numerical evidence of its superior performance has been published elsewhere \cite{Chancellor19a}, and an experimental study of it's comparative performance to one-hot is forthcoming \cite{Chen20a}. This technique has also been used in solving quantum field theory problems using quantum annealers \cite{Abel20a}.

\begin{figure}
\includegraphics[width=7 cm]{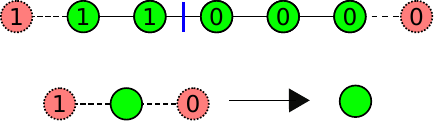}
\caption{\label{fig:domain_wall_cartoon} Top: Encoding of a discrete variable as a domain wall position, the domain wall is depicted in blue, real qubits in green and `virtual' qubits which are fixed are depicted in pink with dotted borders. Bottom: A binary variable in the domain wall encoding reduces to the standard qubit representation.}
\end{figure}

In the interest of making the current paper self-contained, I review the basics of the domain wall encoding and some of it's key features. As Fig.~\ref{fig:domain_wall_cartoon}(top), shows, the domain wall encoding is produced by creating a linearly connected chain of $n-1$ qubit with frustrating fields on each end such that there are $n$ total possible domain wall locations, including frustrating the the fields at either end, which can be thought of as couplings between the terminal qubits and ``virtual'' qubits which are constrained to take either the $1$ or $0$ value. As was discussed in detail in \cite{Chancellor19a}, any interaction between two discrete variables can be realized using two body couplings between the qubits in the domain wall encoding, and arbitrary penalties can be realized by putting fields (single body terms) on the chain. Moreover, the domain wall encoding of a binary variable simply reduced to a normal qubit representation, as depicted in Fig.~\ref{fig:domain_wall_cartoon}(bottom).

I am interested in simple couplings which force frustration in the planted solution problem if the domain wall variable takes one of its soft values, while simultaneously avoiding the need for minor embedding. To do this, I place a single ferromagnetic coupler between the qubits encoding the discrete variables and the other qubits at each end of the soft region. For the additional energy penalties on the chain, I make use of the fact that a single (non-extreme) value of the discrete variable can be penalized using a term of the form $\delta_i=\frac{1}{2}(Z_i-Z_{i-1})$. For the extreme values can be penalized in the same way, but omitting terms which correspond to virtual qubits. This method is described in more detail in \cite{Chancellor19a}, and software for realizing these encodings can be found at \cite{domain_wall_code}.

\bibliography{bibLibrary}  

\end{document}